\documentclass{aa}  
\usepackage{natbib}
\usepackage{graphicx}
\usepackage{txfonts}

\begin{document}

   \title{The undetectable fraction of core-collapse supernovae in luminous infrared galaxies}

   \subtitle{}

   \author{I. Mäntynen\inst{1}\fnmsep\thanks{Corresponding author; \email{iamant@utu.fi}}
          \and
          E. Kankare
          \inst{1}
          \and
          S. Mattila
          \inst{1, 2}
          \and
          A. Efstathiou
          \inst{2}
          \and
          S. D. Ryder
          \inst{3,4}
          \and
          T. M. Reynolds
          \inst{1,5,6}
          \and
          C. Vassallo
          \inst{1}
          \and
          P. Väisänen
          \inst{7,8}
          }

   \institute{Tuorla Observatory, Department of Physics and Astronomy, University of Turku, 20014 Turku, Finland
    \and
    School of Sciences, European University Cyprus, Diogenes street, Engomi, 1516 Nicosia, Cyprus
    \and
    School of Mathematical and Physical Sciences, Macquarie University, Sydney, NSW 2109, Australia
    \and
    Astrophysics and Space Technologies Research Centre, Macquarie University, Sydney, NSW 2109, Australia
    \and
    Cosmic Dawn Center (DAWN)
    \and
    Niels Bohr Institute, University of Copenhagen, Jagtvej 128, 2200 København N, Denmark
    \and
    Finnish Centre for Astronomy with ESO (FINCA), University of Turku, 20014 Turku, Finland
    \and
    South African Astronomical Observatory, P.O. Box 9, Observatory, 7935 Cape Town, South Africa
             }

   \date{Received September 15, 1996; accepted March 16, 1997}

  \abstract 
   {A large fraction of core-collapse supernovae (\mbox{CCSNe}) in luminous infrared galaxies (LIRGs) remain undetected due to extremely high line-of-sight host galaxy dust extinction, and strong contrast between the SN and the galaxy background in the central regions of LIRGs, where the star formation is concentrated. This fraction of undetected CCSNe, unaccounted for by typical extinction corrections, is an important factor in determining CCSN rates, in particular at redshifts $z \gtrsim 1$, where LIRGs dominate the cosmic star formation.}
   {Our aim is to derive a robust estimate for the undetected fraction of CCSNe in LIRGs in the local Universe. Our study is based on the \textit{K}-band multi-epoch SUNBIRD survey data set of a sample of eight LIRGs using the Gemini-North Telescope with the ALTAIR/NIRI laser guide star adaptive optics system.}
   {We used simulated SNe and a standard image subtraction method to determine limiting detection magnitudes for the data set. Subsequently, we used a Monte Carlo method to combine the limiting magnitudes with the survey cadence, and an adopted distribution of CCSN subtypes and their light curve evolution to determine SN detection probabilities. Lastly, we combined these probabilities with the intrinsic CCSN rates of the sample galaxies estimated based on their detailed radiative transfer modeling to derive the fraction of undetectable CCSNe in local LIRGs.}
   {For high angular resolution near-infrared surveys, we find an undetectable fraction of $66.0^{+8.6}_{-14.6}$\%, assuming that CCSNe with host extinctions up to $A_V$ = 16 mag are detectable, corresponding to the most obscured CCSN discovered in our data set. Alternatively, assuming a host extinction limit of $A_V$ = 3 mag, corresponding to typical optical surveys, we find an undetectable CCSN fraction of $89.7^{+2.6}_{-4.4}$\%.}
   {}

   \keywords{supernovae: general -- galaxies: star formation -- dust, extinction
               }

   \maketitle

\section{Introduction}
\label{introduction}

    Core-collapse supernovae (\mbox{CCSN}) are the terminal explosions at the end of the lifetime of massive stars above $\sim$8 M$_{\odot}$ (e.g., \citealt{CCSN_prog_mass}). Due to the relatively short life cycles of massive stars (a few to a few tens of Myr), CCSNe trace the star formation rate (\mbox{SFR}). This offers an approach to estimate the SFR independently from the most commonly used galaxy luminosity based methods. There is some disagreement about whether the observed CCSN rates match the theoretical CCSN rates based on SFR estimates. \cite{horiuchi2011} suggested that up to $z \approx$ 0.4 the predicted CCSN rate is a factor of $\sim$2 higher than the observed CCSN rate. \cite{cappellaro2015} concurred with this claim at 0.0 < \textit{z} < 1.0, while \cite{botticella2017} found disagreements between the observed and the predicted CCSN rates at 0.15 < \textit{z} < 0.35. The variety of methods of estimating SFRs might explain some of the disagreements (see e.g. \citealt{gruppioni2020}). For example, \cite{botticella2012} found that their SFRs estimated from H$\alpha$ and far-ultraviolet luminosities resulted in different CCSN rates between the two methods. \cite{mattila2012} concluded that the likely reason for the discrepancy between the CCSN rate and SFR estimates is a large population of CCSNe that remain undetected due to extreme host extinctions. Their estimate for this additional missing fraction of CCSNe ranges from 19\% at $z=0$ to 38\% at $z\approx$1.2. 
    \cite{melinder2012} and \cite{dahlen2012} used results of \cite{mattila2012} to correct also for the effects of the missing fraction, and found that theoretical and observed CCSN rates matched within errors at 0.1 < \textit{z} < 1.0 and 0.4 < \textit{z} < 1.0 respectively. However, the results of \cite{strolger2015} suggested that the observed and expected CCSN rates were in a reasonable agreement at 0.1 < \textit{z} < 2.5.

    Luminous infrared galaxies (LIRGs) and ultraluminous infrared galaxies (\mbox{ULIRGs}) are defined as galaxies with $ 10^{11}L_{\odot} < L_{\mathrm{IR}} < 10^{12}L_{\odot}$ and $ 10^{12}L_{\odot} < L_{\mathrm{IR}} < 10^{13}L_{\odot}$, respectively, which have enhanced SFRs (e.g. IC 883 with a high SFR of roughly 233 $M_{\odot}$ yr$^{-1}$; \citealt{host_extinction_paper}). The source of the high infrared luminosity can be an active galactic nucleus (\mbox{AGN}), star formation or both. They often show signs of past interaction, or are actively merging. The primary source of $L_{\mathrm{IR}}$ in LIRGs with enhanced SFRs is the reprocessing of light from young hot stars via dust and gas \citep{Kennicutt1998,torres2021}.
    Since the star formation in LIRGs is concentrated in the nuclear and circumnuclear regions in the scale of $\sim$0.1 to $\sim$1 kpc \citep{soifer2001}, a considerable fraction of SNe are not discovered due to extinction and insufficient spatial resolution of the observations \citep{sunbird}. The differences between observed and predicted CCSN rates are likely to only increase at larger redshifts, where LIRGs start to dominate star formation, with increasing contribution as redshift increases \citep{lefloch2005,magnelli2011}. At \textit{z} = 1 the contribution of LIRGs is $\sim$50\% to the SFR (e.g. \citealt{magnelli2009}).

    CCSN rate studies typically correct their observed rates with extinction correction models (e.g. \citealt{riello2005}). However, a fraction of CCSNe can have very high localised extinctions that these extinction correction models do not take sufficiently into account. This population of CCSNe with extreme host extinctions that remain undiscoverable, and unaccounted for after CCSN rate extinction corrections, results in an additional fraction of CCSNe that are missed by surveys, hereafter referred as the `missing fraction'. Furthermore, we will refer with the term `undetectable fraction' to the combined fraction of SNe that would remain undiscovered due to belonging either to the population with sufficiently high host extinctions described by the extinction correction model or to the population described by the missing fraction.
    
    \cite{mattila2012} estimated that as much as $83^{+9}_{-15}\%$ of CCSNe in LIRGs are missed by optical surveys due to extreme host extinctions. This estimate is based on a number of CCSNe found in one nearby LIRG system (Arp 299) by several surveys over 14 years. \cite{miluzio2013} found that in ground-based seeing \textit{K}-band observations, where the extinction is $\sim$10\% of the optical extinction, the percentage of CCSNe missed in LIRGs can be as high as $\sim60\%$ to $75\%$. Their estimate was based on artificial star experiments on their sample galaxies. This was in general agreement with a previous result by \cite{mannucci2007}, which estimated that $\sim$30\% of CCSNe are missed at \textit{z} $\approx~$ 1, since 50\% of SFR is in LIRGs at that redshift. A survey with the Spitzer Space Telescope at 3.6 $\mu$m by \cite{fox2021} monitored a sample of (U)LIRGs for 2 years, and detected 9 CCSNe out of the expected intrinsic number of 51.9 CCSNe in the 32 (U)LIRGs at $<$150 Mpc within which the survey was found to be sufficiently sensitive; this suggests that $\sim$83\% of CCSNe remained undetected.
    
    An excellent alternative to space-based surveys are monitoring observations with adaptive optics (AO) at near-IR wavelengths from 8 metre-class ground-based telescopes, as shown by multiple CCSN discoveries in LIRGs \citep{AO_SN1,iras17138_SN,ugc_SN,sunbird}. 
    The distribution of all optical or infrared detected CCSNe in LIRGs was presented by \citeauthor{sunbird} (2018, see their figure 11) with the projected nuclear offsets and discovery methods reported. While optical surveys found most of the CCSNe in the sample at nuclear distances of >4 kpc, non-AO and AO NIR surveys dominate closer to the nucleus when taking into account the smaller sample size. The data also suggests that performing a survey in less dust-obscured NIR bands alone is not enough due to the poorer spatial resolution of non-AO NIR surveys compared to the high-resolution methods, and a combination of AO and NIR bands are required to effectively detect CCSNe in the nuclear regions of LIRGs. Therefore, the usage of near-IR bands compared to optical bands has the most important influence on the detectability of obscured CCSNe in LIRGs, with the high-resolution methods providing a crucial addition for the surveys.
    Thus, accurate estimates for the percentage of CCSNe that remain undetected by CCSN rate surveys can be estimated with improved accuracy using this high-resolution method.

    Deep IR observations with the James Webb Space Telescope (JWST) have recently started producing CCSN discoveries at redshifts beyond z = 3 \citep{decoursey} whereas Euclid and Roman Space Telescope will be increasing substantially the CCSN rate statistics at $z \gtrsim 1$. Reliable corrections for the CCSNe remaining undetected by these surveys due to very high extinctions in dusty galaxies across the redshift (e.g. \cite{mattila2012} will allow detailed comparisons with the rates expected from the cosmic SFRs.
    
    CCSNe are classically divided into subtypes based on their lightcurves and spectral features \citep{Filippenko1997}. This work considers CCSN subtypes II, IIn, IIb, Ib, Ic and SN 1987A-like. Type IIb, Ib and Ic are called stripped envelope SNe and exhibit spectral features indicative of increased envelope stripping, showing some hydrogen, helium but no hydrogen, and neither hydrogen nor helium, respectively. Type II refers to hydrogen-rich SNe, which are sometimes further classified to Type II-P and II-L based on their light curve evolution. Type IIn SNe are hydrogen-rich, but exhibit narrow spectral lines ($\lesssim$1000 km s$^{-1}$). Finally, SN 1987A-like events are peculiar hydrogen-rich CCSNe that resemble SN 1987A with a slowly rising optical light curve.

    Our method of analysing the data (Sect. \ref{data_set}) was the following. Algorithmic image subtraction (Sect. \ref{autom_image_sub}) was used as a method to detect SNe in the data set. The limiting magnitudes (Sect. \ref{limiting_magnitudes}) were determined for CCSN detections via injection of artificial SNe and automated detection of these sources. Subsequently, a Monte Carlo method (Sect. \ref{monte_carlo_method}) was used for collating effects of different CCSN subtypes, peak magnitude variation, evolution time scale variability of CCSNe, host extinction, and observational cadence to produce CCSN detection probabilities for the data set. The intrinsic CCSN rates were estimated for the data set using modeling of the spectral energy distribution (SED) of the sample LIRGs, where radiative transfer models for different components of the galaxy were fitted to the SEDs. The intrinsic CCSN rates were then compared with the observed rates to derive the missing fraction estimate, and the adopted host extinction model was used to derive the survey depth dependent fraction of CCSNe that would remain undiscovered. These two effects were combined to derive the total undetectable fraction of CCSNe in LIRGs, and is reported in Sect. \ref{results}. Conclusions are provided in Sect. \ref{conclusions}.

\section{Data set}
\label{data_set}

    The data set analysed in this study was obtained by the Supernovae UNmasked By Infra-Red Detection (SUNBIRD) collaboration \citep{sunbird} with the aim to discover and study CCSNe in high-SFR LIRGs. The data consists of 67 epochs of eight different LIRGs in total, spanning from March 2007 to August 2012, see Table \ref{data set_table}. The data were obtained using the Gemini-North Telescope with the Near-InfraRed Imager (NIRI; \citealt{niri}) and the ALTAIR laser guide star AO system  $(0.022''$ pixel$^{-1})$ in \textit{K}-band. The total on-source exposure times of the analysed images were typically 9 $\times$ 30 s; however, in a few cases the number of co-adds varied from a minimum of four to a maximum of 18. Each 30 s image was obtained in individual exposures of one to ten seconds depending on the brightness of the galaxy nucleus. No separate frames were taken for sky background subtraction, instead the sky images were created from the dithered 30 s images of the target. The data reduction was carried out with the external \texttt{gemini} package in \texttt{IRAF} \citep{IRAF} to perform the typical steps of flat field correction, sky subtraction, and co-addition of the processed images. For each galaxy, all the ALTAIR/NIRI images were aligned to match the epoch with the best seeing using the \texttt{IRAF} tasks \texttt{geomap} and \texttt{geotran}.

    The photometric calibration was based on the Two Micron All Sky Survey (2MASS; \citealt{skrutskie2006}). Due to the small 22" $\times$ 22" field of view (FOV) of the ALTAIR/NIRI setup, there are no isolated 2MASS point sources in the target frames. Magnitudes of bootstrapped sequence stars in the fields were adopted for IRAS 17138-1017, IC 883, Arp 299-A, and Arp 299-B from the analyses of \cite{iras17138_SN,ugc_SN,Arp299_SNs}, previously obtained for the studies of CCSNe discovered in these LIRGs. For the calibration of the images of the other LIRGs, we made use of \textit{Ks}-band images of the galaxy fields obtained using the Nordic Optical Telescope (NOT; \citealt{NOT_telescope}) with the NOT near-infrared Camera and spectrograph (NOTCam) instrument. The NOTCam images were reduced using the external \texttt{notcam} package in \texttt{IRAF}, which included flat field correction, sky subtraction, and co-adding individual frames for higher signal-to-noise ratio. With the NOTCam FOV of 4$\arcmin$ $\times$ 4$\arcmin$ typically several 2MASS point sources were found in the reduced images to calibrate their zeropoints. Subsequently, common point sources in both the NOTCam and the ALTAIR/NIRI images were identified and their magnitudes were measured to use them for the calibration of the ALTAIR/NIRI data of IRAS 16516-0948 and IRAS 17578-0400. The aperture photometry was carried out using the \texttt{Source Extractor} program \citep{SExtractor}. In the case of two galaxies, CGCG 049-057 and MCG+08-11-002, there were no point sources in the ALTAIR/NIRI images. In these cases the whole galaxy was measured with a largest possible aperture for the ALTAIR/NIRI images, and calibrated accordingly with an identical aperture using the NOTCam images. In these two galaxies the aperture photometry was performed with \texttt{IRAF} due to the poor performance of \texttt{Source Extractor} with extended objects.

    The point spread function (PSF) of the AO images includes a narrow core and extended wings. The exact sizes of these and the Strehl ratio depends on observational conditions, for example seeing, brightness of the natural guide star and its distance from the target. In our data set, the median measured full width at half maximum (FWHM) of the PSF core was six pixels (corresponding to 0.13"), the standard deviation was 3.1 pixels, and the 95th percentile was at 9.4 pixels. Performing aperture photometry with conventional aperture sizes can miss a significant portion of the flux from the extended wings. We measured the effect of this by performing aperture photometry on a field star from an AO image with a progressively larger aperture until the measured brightness of the star remained constant. The resulting aperture correction was the difference between this end magnitude and the magnitude at the aperture size used in the photometry. The bootstrapped magnitudes from the NOTCam images were then corrected by this aperture correction. 

    Five CCSNe have been previously reported based on the ALTAIR/NIRI data set: SN 2010cu (Type IIP, $A^{\mathrm{host}}_V$ = 0 mag) and SN 2011hi (Type IIP, $A^{\mathrm{host}}_V$ = 6 mag) in IC 883 \citep{ugc_SN}, SN 2010O (Type Ib, $A^{\mathrm{host}}_V$ = 1.9 mag) and SN 2010P (Type IIb, $A^{\mathrm{host}}_V$ = 7 mag) in Arp 299-A/B \citep{Arp299_SNs} and SN 2008cs (Type IIn, $A^{\mathrm{host}}_V$ = 16 mag) in IRAS 17138-1017 \citep{iras17138_SN}. No new previously unreported CCSN candidates were discovered in the data set during our analysis. The discovered number of CCSNe was compared to the expected number of events in our analysis.

\begin{table*}[h]
\caption{Summary of the data set with the number of epochs, \textit{N}, survey coverage in days, \textit{t}, estimated SFR value, SFR, starburst age, $t_{\mathrm{SB}}$, intrinsic rate of CCSNe, CCSN rate, AGN contribution to the luminosity, $L_{\mathrm{AGN}}$, luminosity distance, \textit{D}, and Galactic line-of-sight extinction, $A^{\mathrm{Gal}}_K$. The luminosity distances are corrected for the influence of the Virgo Cluster and the Great Attractor infall, and assumed $H_0 = 70$ km s$^{-1}$ Mpc$^{-1}$, $\Omega_{\mathrm{M}}=0.3$, $\Omega_{\Lambda}=0.7$. The SFR, CCSN rate, and $L_{\mathrm{AGN}}$ values for IC 883, Arp 299-A/B, and IRAS 17138-1017 are adopted from \cite{host_extinction_paper}}.
    \centering
    \begin{tabular}{ccccccccc}
    \hline
    \hline
        Galaxy & \textit{N} & \textit{t} & SFR & $t_{\mathrm{SB}}$& CCSN rate & $L_{\mathrm{AGN}}$ & \textit{D} & $A^{\mathrm{Gal}}_K$\\
          &  & (d) & ($M_{\odot}$ yr$^{-1}$) & (Myr) & (SN yr$^{-1}$) & (\%) & (Mpc) & (mag) \\
    \hline
        IC 883 & 11 & 1386 & $233^{+21}_{-14}$ & $38^{+1}_{-1}$ & 2.1$^{+0.1}_{-0.1}$ & 45$^{+13}_{-20}$ & 106.1 & 0.004 \\
        MCG+08-11-002 & 4 & 504 & $125^{+7}_{-10}$ & $36^{+1}_{-1}$ & 1.2$^{+0.1}_{-0.1}$ & 28$^{+8}_{-4}$ & 82.6 & 0.091 \\
        IRAS 17578-0400 & 9 & 1562 & $107^{+4}_{-4}$ & $36^{+1}_{-1}$ & 1.1$^{+0.1}_{-0.1}$ & 0 & 67.2 & 0.431 \\
        Arp 299-A & 8 & 1318 & $59^{+9}_{-16}$ & $16^{+2}_{-4}$ & 0.9$^{+0.1}_{-0.2}$ & 0 & 49.9 & 0.005 \\
        IRAS 17138-1017 & 9 & 1564 & $69^{+37}_{-4}$ & $28^{+7}_{-1}$ & 0.8$^{+0.1}_{-0.1}$ & 31$^{+20}_{-4}$ & 82.2 & 0.207 \\
        CGCG 049-057  & 8 & 1566& $48^{+15}_{-3}$ & $20^{+5}_{-1}$ & 0.6$^{+0.1}_{-0.1}$ & 10$^{+2}_{-10}$ & 64.7 & 0.012 \\
        Arp 299-B & 11 & 1790 & $40^{+9}_{-16}$ & $16^{+2}_{-4}$ & 0.6$^{+0.1}_{-0.2}$ & 0 & 49.9 & 0.005 \\
        IRAS 16516-0948 & 7 & 1563 & $50^{+4}_{-4}$ & $27^{+1}_{-1}$ & 0.5$^{+0.1}_{-0.1}$ & 37$^{+4}_{-8}$ & 105.5 & 0.226 \\
    \hline
    \end{tabular}
    \label{data set_table}
\end{table*}

\section{Method}
\label{method}

    We begin by summarizing the overall method. First, we developed an automated process for the template subtraction of the survey images, which was carried out for the data set. Second, we derived spatially variable limiting magnitudes for each epoch of each galaxy. Third, we derived detection probabilities of individual simulated SNe for each galaxy in the data set via a Monte Carlo simulation. Finally, the detection probabilities were combined with intrinsic CCSN rates estimated with radiative transfer modeling of the SEDs  of the sample galaxies to derive the missing fraction of CCSNe in LIRGs. The undetectable fractions of CCSNe were derived combining the results from the host extinction model and missing fraction estimate. Details of these individual steps are provided in the following subsections.

\subsection{Automatic image subtraction}
\label{autom_image_sub}

\begin{figure*}[hbt]
    \begin{center}
        \includegraphics[width=1.\textwidth]{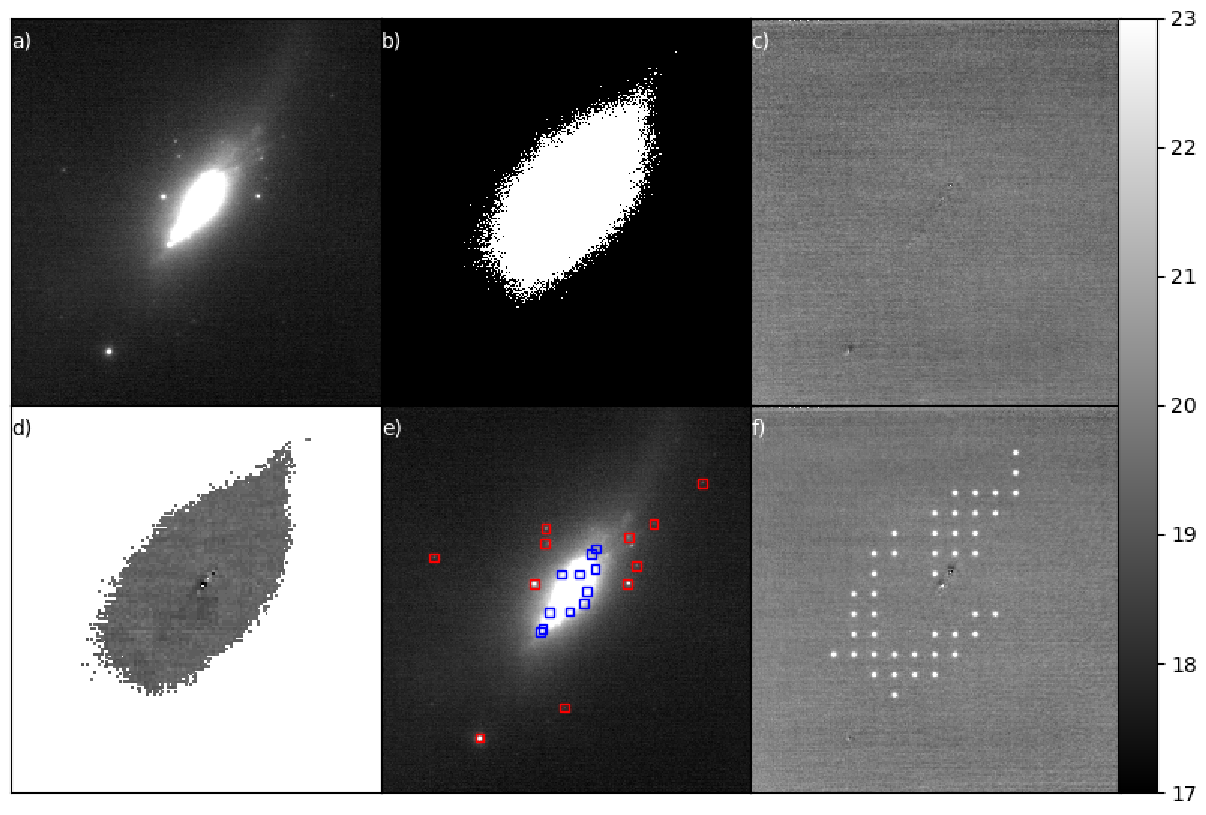}
        \end{center}
    \caption{a) A 22" $\times$ 22" median stacked image of the sample galaxy IC 883. b) Galaxy map constructed from the median stacked image. c) An example of a template subtraction. d) Limiting magnitudes of IC 883 with the scale on the right. The two white pixels in the center of the galaxy represent areas where persistent template subtraction artefacts prevented analysis; further details are provided in Sect. \ref{limiting_magnitudes}. e) Example stamp positions for four sets of stamp lists. Red squares denote the non-varied and blue squares denote the varied stamps in the stamp lists. The positions of both sets would be changed in a subsequent run to ensure that the whole galaxy was covered. f) An example of the limiting magnitude iteration process. Visible point sources are simulated SNe at 16.8 mag, which are positioned on the galaxy map, excluding the stamp areas, separated by 7 $\times$ FWHM = 42 pixels.}
    \label{example_fits}
\end{figure*}

    We used the method described by \cite{sum_method} to create galaxy region maps for each of the sample LIRGs in the ALTAIR/NIRI data. First, we created a median stacked image for every sample galaxy, where each pixel value is the median value of that pixel from all the aligned epochs of the images. Then a background level was subtracted from the median image. This background level was selected as an area far from the galaxy and field stars. Finally, the pixel values were summed from the lowest to the highest. The ranked pixels are considered to be part of the background until the sum reaches zero, and the not yet included pixels belong to the galaxy. A few hundred pixels wide regions at the edges of the images with a lower signal-to-noise ratio were excluded from the process. Field stars were manually removed from the galaxy maps. An example of a median stacked image and the corresponding galaxy map is shown in Fig. \ref{example_fits}. The exact location of the background sample area does have some effect on the galaxy maps (i.e. in the extent of the outer edges of the galaxy). However, the effect of this on our method is minimal, since the expected CCSN rate at the outer edges of the galaxy is low.

    Image subtraction is a basic method used by SN search programs. Sources in an image that remain constant between epochs (e.g. stars and galaxies) disappear in a good quality subtraction. Conversely, transient sources (e.g. sufficiently bright CCSNe) remain detectable in the subtracted image if the source brightness has evolved sufficiently between the epochs or the source is not present in one of the images. However, in practice, the image subtraction methods can leave residuals in the subtracted image, in particular to the brightest regions of the field. Furthermore, the noise in the image is increased by a factor of $\sqrt{2}$ in the process. 
    The optimised subtraction parameters were obtained for each image in the data set. For each image the subtraction template image was selected based on having the most similar image quality and at least 200 days difference in time between the epochs of the images.
    
    The image subtraction process in our analysis was carried out using a slightly modified version of the \texttt{ISIS2.2} package \citep{ISIS2,ISIS}. The user provides the software two aligned images, with a similar PSF for best results, and enables the software to select "stamp" regions automatically or provides them externally for the code. These stamps consist of two co-centric squares with inner and outer areas. The inner areas are used to generate a convolution kernel to convolve the image with the narrower PSF to match the image with the broader PSF. The outer area is used to measure the background of the image.
    After calculating the convolution kernel, the software performs the image convolution, scales the fluxes to match, and subtracts the images pixel by pixel. The quality of the image subtraction result is highly dependent on some of the adopted parameters (e.g. \citealt{melinder2012}). In order to reduce the subjective bias in this process, we developed an automatic algorithm to optimise the parameter selection.

    Our algorithm optimises three key image subtraction parameters for \texttt{ISIS2.2}, assuming that they are largely independent: the kernel size, the size of the background fit region and the degree of the kernel fit. The degree of the kernel fit dictates how much the kernel is allowed to vary across the image. Too high of a degree compared to the number of stamps provided can cause overfitting. The algorithm starts with a set of initial guesses for the values for kernel and stamp size provided by the user. The program then performs a subtraction using all the combinations of the initial guesses. It then chooses the best combination of the kernel and stamp sizes. The quality of the subtraction is measured as the variance of pixel values across the area included in the galaxy map, excluding the stamp regions. A new set of values for the parameters are then generated above and below the previously selected best values. The program then performs subtractions using all the combinations of these new values, and again chooses the best ones. This iterative process continues until no improvements are made anymore, or the upper or lower bounds of the parameters set by the user are reached. One caveat of this approach would be that it might converge to some local minima, and miss the optimal parameter values. However, according to our tests, the quality of the subtraction evolves smoothly as a function of these variables, and thus this is not a major concern. After kernel and stamp sizes are set, the algorithm tests which degree of kernel variation is optimal.
    
    Furthermore, our algorithm varies the selection of the used stamps from an initial list provided by the user (see Fig. \ref{example_fits}). If a SN occurs within a stamp location used to derive the convolution kernel by the image subtraction package, the SN might disappear or become fainter in the process. \texttt{ISIS2.2} includes an automatic step that should reject such stamps; however, this is uncertain if the SN is faint compared to the surrounding pixels and close to the detection threshold. Therefore, the algorithm is given two sets of stamps: one set that is fixed during a run and a second set which is varied during the optimisation process. The results of the process are two lists of optimal stamp locations and corresponding image subtraction parameter values, where the varied stamp locations do not overlap between the two lists. In an optimal scenario two varied stamp lists is enough to cover the entire galaxy; however, in most cases one to two additional varied stamp lists are required to ensure high-quality subtractions over the whole galaxy when the results are combined. An example subtraction is shown in Fig. \ref{example_fits}.

\subsection{Limiting magnitudes}
\label{limiting_magnitudes}

    Noise in the subtracted images and image subtraction residuals affect the detection probabilities of SNe. To take this into account, we developed a program to simulate artificial SNe in sequences of different magnitudes on the ALTAIR/NIRI images with the aim to recover them after the image subtraction process to derive the detection thresholds. This results in spatially variable limiting magnitudes for each epoch. An example of a final product of this method is shown in Fig. \ref{example_fits}.

    The artificial SNe were created with the \texttt{IRAF} routine \texttt{mkobjects}. Simulated SNe have a Moffat profile with a Moffat parameter of 2.5, which is typical for seeing limited profiles, and a seeing value that matches the FWHM of the narrow core PSF of point sources in the analysed image. This approach was chosen since most of the sample galaxies in the ALTAIR/NIRI data did not have bright point sources in the FOV and two of the LIRGs (MCG+08-11-002 and CGCG 049-057) did not have any point sources from which to determine an accurate description of the full AO PSF including the faint and extended wings. Thus for consistency across the data set we used the Moffat profile for the whole data set. In order to estimate the effect of this, we repeated the limiting magnitude procedure for the LIRG IRAS 17578-0400 with the usage of a PSF of a bright star in the FOV (including the AO PSF wings) instead of the Moffat profile. The resulting limiting magnitude differences from those with the Moffat profile were on average less than the 0.2 mag step used to derive the detection threshold. Therefore, the usage of the Moffat profile to describe the dominant narrow PSF core seems sufficient for our analysis.
    
    The SNe are simulated on the pixels based on the galaxy map; however, limited to a grid size that corresponds to the FWHM of the narrow PSF core of point sources in the image. For example, for a typical image with a FWHM $\approx$ 5 pixels, the artificial SNe are simulated on the galaxy map covered area in a grid where the sources are separated by 5 pixels in both x and y direction. However, if all these artificial SNe would be created in the image simultaneously, their detection as individual sources would be impossible due to the blending of the sources; therefore, the complete grid of simulated SNe has to be divided into sub-grid parts. Based on our tests, a separation of 7 times the image FWHM between each SN was sufficient to prevent any notable overlap effects. Therefore, the initial grid is divided into 64 sub-grids where the artificial SNe are separated by $\geq$7 $\times$ FWHM. An example of a sub-grid is presented in Fig. \ref{example_fits}. 
    
    An iterative process is then carried out for these 64 sub-grids consecutively. The automatic process consists of creating artificial SNe with \texttt{IRAF}, carrying out the image subtraction with \texttt{ISIS2.2}, detecting the SNe with \texttt{Source Extractor}, and iterating the process making the SNe fainter until none are detected. Areas within 10 $\times$ FWHM of known real SNe were excluded from the analysis in epochs where the events were detectable. This was done to avoid any pre-existing SN from affecting the detection of nearby simulated SNe. We note that the probability of two simultaneous SNe with overlapping locations is extremely low and the excluded regions were small.

     Simulated SNe were detected from the subtracted images using \texttt{Source Extractor} with the following parameters: minimum pixel area of 9 and a detection threshold of 5$\sigma$ above background. The minimum pixel area was chosen to strike a balance between weeding out most image subtraction artefacts, while having a minimal effect on the detection of dim artificial SNe. The 5$\sigma$ level was chosen conservatively, and result in a threshold where the SNe are sufficiently distinguishable by eye. Occasionally noise effects shifted the location of the simulated SNe approximated by \texttt{Source Extractor} in the subtracted image. To prevent this effect from causing false non-detections, a simulated SN was counted as a detection if a source was detected within four pixels of a simulated SN position. The range of four pixels was found by testing to be sufficient, while not causing false positives.
     Since \texttt{Source Extractor} does not perform PSF fitting of the sources, it is more challenging to avoid identifying artefacts as detections compared to PSF fitting based methods. These false detections were confirmed to be artefacts rather than real detections by inspecting their light profiles. Their profiles were unnaturally narrow and asymmetric to be real detections. The locations with an artefact in the subtracted image were not analysed with simulated SNe. The effect of these artefacts was mitigated by performing multiple sets of image subtractions; however, in some unfortunate cases artefacts can appear in the same locations of subtracted images also with different sets of selected stamps. In these few cases those grid points are conservatively considered as non-detectable for the purpose of the analysis. 

    The iterative process to determine the detection threshold magnitudes was the following. Initially, a sub-grid set of artificial SNe with an apparent magnitude of 16 mag were simulated on the image to be analysed. This was considered as a conservative starting magnitude for the simulated sources, which would be clearly detectable in a vast majority of the image subtracted pixels. The image subtraction was then carried out with optimal parameters and stamp selections determined by the algorithm, and the resulting artificial SN detections were recorded. Subsequently, the process was repeated with simulated SNe that are 0.2 mag fainter than before, until either all SNe result in non-detections, or 40 iterations have been carried out. Following this, another spatially shifted sub-grid of SN positions is analysed, and the process continues until the full set of 64 sub-grids is complete. Finally, if there were any simulated SNe that were undetected at 16 mag, the complete process was repeated for those locations starting at 14 mag. The 14 magnitude limit for the brightest simulated SNe was chosen to roughly match the brightest expected CCSNe in the nearest galaxy of the data set. 
    For those artificial SN positions that had a detection threshold magnitude determined more than once in a set of stamp lists, the faintest magnitude was chosen as the threshold. 
    The limiting magnitudes were determined for each epoch in the data set, and each pixel in the galaxy maps. If the two epochs used in the subtraction had very similar FWHM values, the same detection limits were adopted for both images.

\subsection{Monte Carlo method}
\label{monte_carlo_method}

    \cite{miluzio2013} used a Monte Carlo method to estimate SN detection probabilities in their data set. This involved simulating SNe in different regions of their galaxy sample, with varying luminosities and using \textit{K}-band light curve models for the different SN subtypes. Depending on the explosion dates, extinction parameters and limiting magnitudes, the individual SNe were either counted as detections or missed events. Our approach is similar to their work, with the focus on AO searches for CCSNe in LIRGs extending over more years. 

    The individual components required for the method were the following: detection thresholds for the faintest detectable SNe in each pixel in each epoch of the data set, templates for the light curve evolution for each CCSN subtype and limits within which they vary, a model for the host extinction, the survey cadence, the adopted relative rates of CCSN subtypes, and finally the intrinsic total CCSN rates for the data set. Each run of the Monte Carlo program simulated $10^7$ SNe. The following procedure was used. Time resolution for the simulation was set to one day and spatial resolution to one pixel in the galaxy maps. First a CCSN subtype is chosen, then a random day is chosen within the typical detection window  (i.e. control time, see Sect. \ref{control_times}) for that subtype of chosen galaxy as the explosion date. Then a pixel in the galaxy map is randomly chosen as the SN location based on the probability to host a CCSN. This probability is assumed to be directly proportional to the ratio of the flux within a radius of five pixels around that position and the total flux of the LIRG in \textit{K}-band. The total flux is the measured sum of the pixel values within the galaxy map area in the median stack of all epochs of the galaxy. Next the CCSN subtype template light curve is generated, and its absolute brightness is determined based on the time to the next observation epoch. Finally the distance modulus is added to the absolute magnitude and the galactic and host extinction effects are included. The resulting SN apparent magnitude is then compared against the limiting magnitude of that pixel in the epoch where the SN would be first observable. If the SN is brighter than the limit, it is counted as a detection.

    \subsubsection{Template light curves}
    \label{template_light_curves}

    CCSN subtypes included in the simulation are: II, IIn, IIb, Ib, Ic and SN 1987A-like. The relative rates of these CCSN subtypes were adopted from the observed volume-limited results of the Lick Observatory Supernova Search (LOSS, \citealt{Li2000}) with CCSN classifications revised by \cite{SN_subtype_rates}. Our work adopts the CCSN magnitudes and their revised subtypes with small modifications; we do not separate Type II SNe into subtypes Type II-P and II-L, as there is some evidence of them being the same population (e.g. \citealt{anderson2014}), the relative rates of Type IIb-peculiar SNe were merged with the Type IIb SNe, and both Type Ic-peculiar and Type Ic-BL SNe were included in the Type Ic SN fraction. This is justified by the low rates combined with the lack of high-quality \textit{K}-band light curves for these rare subtypes. Type Ia SNe were not considered in the analysis due to their low relative rates ($\sim$ 5\%) compared to CCSNe in high SFR galaxies (see e.g. \citealt{AO_SN1}).
    These relative rates assume identical explosion time windows for the CCSNe, which is not the case in the Monte Carlo simulation. The relative rates are slightly adjusted for each galaxy to counteract this effect. For example, if one CCSN subtype has double the time window to explode compared to the rest, its relative rate needs to be doubled so that the number of CCSNe per year for that subtype stays constant. 

    Requirements for the \textit{K}-band SN light curve data sets from the literature to be adopted as templates were the following: smooth evolution with small photometric errors and an adequate cadence, early photometric points to sufficiently cover the light curve rise and peak, and a sufficient late-time coverage. Furthermore, the explosion date of the SN was required to be well constrained. This resulted in only a few candidates for each subtype. Thus, it was opted to adopt for each subtype only the most representative high-quality light curve as the template. Due to the similar \textit{K}-band light curve evolution of the stripped-envelope SNe (types IIb, Ib, Ic) these were represented by the well observed Type IIb SN 2011dh \citep{SN2011dh_lightcurve1,SN2011dh_lightcurve2}. For the Type II template light curve SN 1999em was chosen \citep{SN1999em_lightcurve}. For the Type IIn template light curve SN 1998S was chosen \citep{SN1998S_lightcurve1,SN1998S_lightcurve2}. For the SN 87A-like light curve SN 1987A was chosen \citep{SN1987_lightcurve1,SN1987_lightcurve2}. For these subtypes and the selected template SNe relevant information is listed in Table \ref{sn_models}. 

    The unfiltered peak absolute magnitude distributions for each SN subtype were obtained from the sample of \cite{peak_mags}, corrected with the updated SN classifications by \cite{SN_subtype_rates}. Only CCSNe with unambiguous classifications were included. Since the data by \cite{peak_mags} does not include any estimates or corrections for the host galaxy extinction, we limited the sample selection to SNe with host inclinations of $<75^{\circ}$ to reduce the effect of the extinction in our absolute magnitude distributions by excluding the most edge-on galaxies. This reduced the number of CCSNe used for these distributions from 88 to 69.
    The peak absolute magnitude distributions of \cite{peak_mags} were obtained without a filter, which is best approximated by \textit{R}-band. 
    
    We examined CCSN light curves from the literature and measured their \textit{R-K} colours at both the \textit{R} and \textit{K}-band light curve peaks, see Table \ref{sn_models}. We limited our peak colour sample to CCSNe with the same criteria as used for the template lightcurves, which resulted in four H-poor SNe and four Type II SNe. For Types IIn and SN 87A-like we used only the model SNe themselves due to the poor availability of high quality lightcurves. For SN 1998S we adopted a more recent extinction estimate by \cite{shivvers2015} of $A_V = 0.16$ mag, as opposed to the value of $A_V = 0.68$ mag by \cite{SN1998S_lightcurve1}. The four H-poor SNe included here were: SN 2002ap \citep{borisov2002}, SN 2007gr, SN 2009jf \citep{bianco2014} and SN 2011dh \citep{SN2011dh_lightcurve1,SN2011dh_lightcurve2}. The four Type II SNe included were: SN 1999em \citep{SN1999em_lightcurve}, SN 2007aa, SN 2008in \citep{hicken2017} and SN 2013ej \citep{yuan2016}.
    We applied the yielded colour values to the unfiltered absolute peak magnitude distributions to obtain \textit{K}-band absolute peak magnitude distributions. An alternative approach would have been to derive peak magnitude statistics directly from the \textit{K}-band light curves from the literature, which we deemed as an inferior method due to the poor availability of \textit{K}-band absolute peak magnitudes.

\begin{figure*}[hbt]
    \begin{center}
        \includegraphics[width=1.0\textwidth]{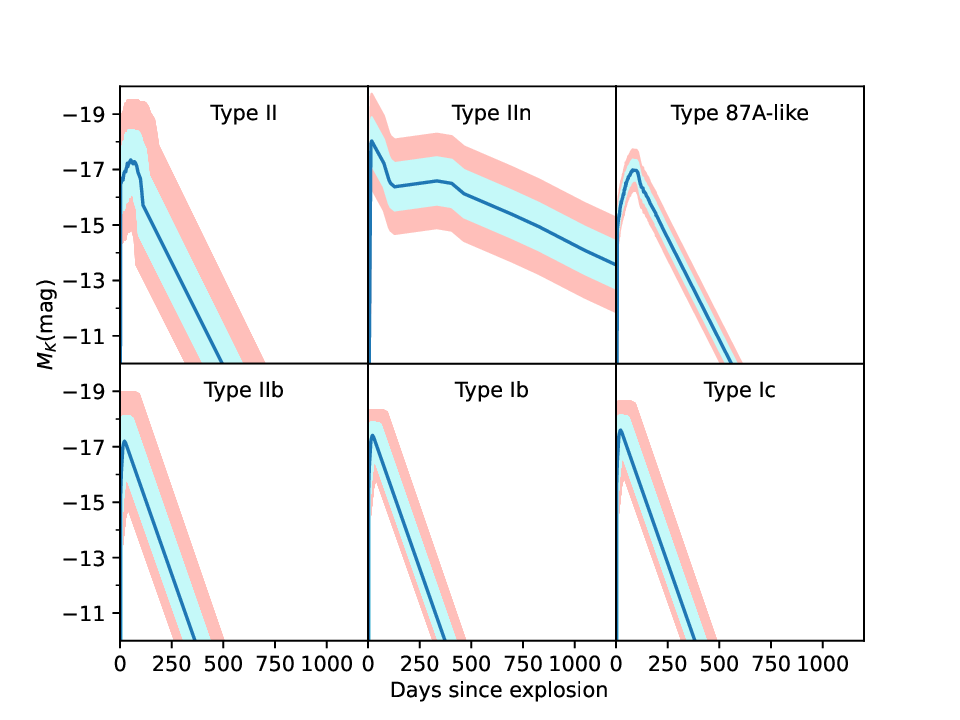}
        \end{center}
    \caption{Possible ranges of the \textit{K}-band template light curves. Dark blue line is the median value and light blue and red areas are the ranges within 1$\sigma$ and 2$\sigma$, respectively. All parameters follow a Gaussian distribution, except for the H-poor templates. Their fastest possible evolution is limited to match the fastest SN in our template sample to avoid unrealistically rapid evolution. All templates are extrapolated with a linear tail phase.}
    \label{light_curves_figure}
\end{figure*}

    For H-poor SNe, the time evolution of the template was multiplied by a factor based on the distribution of light-curve decline rate, $m_{\Delta15}$. This parameter is defined as the difference in magnitude between peak and 15 days after peak \citep{delta_15}. The SNe used here were the same that were used in the peak colour investigation. This time factor is based on the statistics of the $m_{\Delta15}$ values of the sample. The factor is normally distributed, with the limitation that the factor cannot make the SN template evolve faster than the fastest evolving SN of the sample, to avoid unphysically narrow light curve shapes. The probability of this occurring is outside the 2$\sigma$ variability of the time factor; therefore, this does not skew the shape of the distribution in a major way. This factor was only applied to first 35 days of the template. This was done to prevent the late time evolution from being unphysically fast or slow.
    A similar approach was taken with Type II SNe, except instead of $m_{\Delta15}$ we used the optically thick phase duration (OPTd; \citealt{anderson2014}), and applied the factor only to the length of the plateau phase of the light curve. The CCSN subtype templates and ranges for their variability are shown in the Fig. \ref{light_curves_figure} and the adopted time factor parameters are listed in Table 2.

\begin{table*}[h]
\caption{Template light curve parameters. Time factor values are the 2$\sigma$ ranges for the time scale variation of the template. Rates are the revised relative LOSS rates for the CCSN subtypes.}
    \centering
    \begin{tabular}{ccccccc}
    \hline
    \hline
        Type & Template & $M_{K}$ & Time factor & Rate & $M_R - M_K$ \\
            &           &(mag) &       &  (\%)   & (mag) \\
     \hline
        Ib & SN 2011dh & -17.41$\pm$0.45 & 0.65-2.20 & 10.8 & 0.51$\pm$0.26 \\
        Ic & SN 2011dh & -17.60$\pm$0.51 & 0.65-2.20 & 8.6 & 0.51$\pm$0.26 \\
        IIb & SN 2011dh & -17.20$\pm$0.88 & 0.65-2.20 & 10.9 & 0.51$\pm$0.26 \\
        II & SN 1999em & -17.35$\pm$1.07 & 0.83-1.24  & 62.2 & 1.00$\pm$0.07 \\
        87A & SN 1987A & -16.99$\pm$0.37 & 1.00 & 2.9 & 1.08 \\
        IIn & SN 1998S & -17.92$\pm$0.85 & 1.00 & 4.6 & 0.11 \\
    \hline
    \end{tabular}

    \label{sn_models}
\end{table*}

\subsubsection{Host extinction}
\label{host_ext}

\begin{figure}[hbt]
    \begin{center}
        \includegraphics[width=0.5\textwidth]{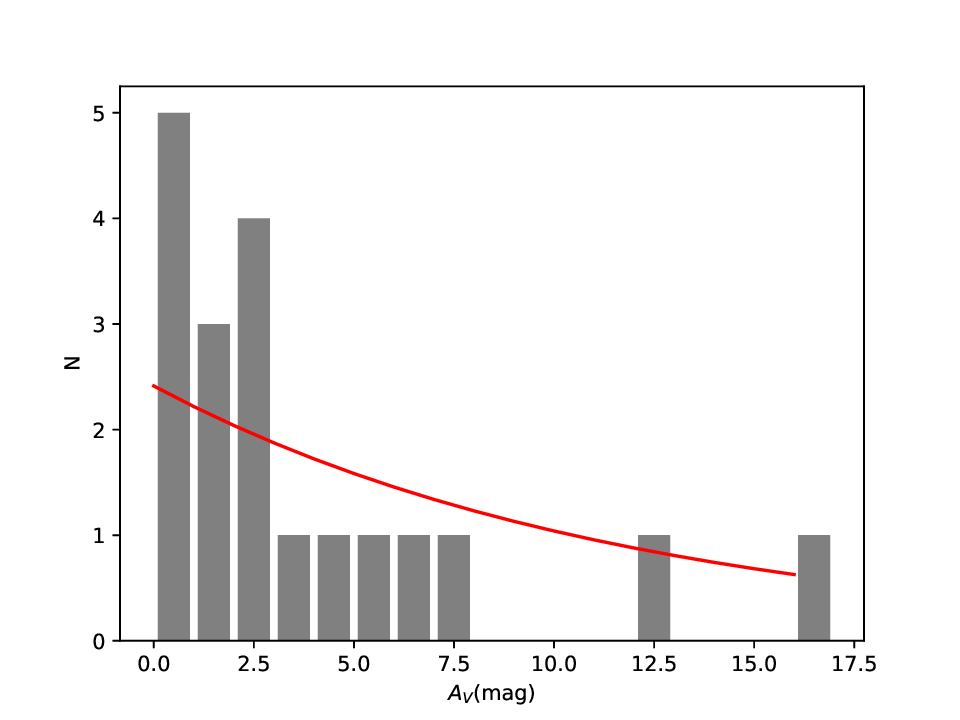}
        \end{center}
    \caption{Host extinction of CCSNe in central regions of LIRGs from \cite{host_extinction_paper} fitted with an exponential function (red curve) resulting in $P(A_V)=0.881e^{-0.084A_V}$.}
    \label{extinction_figure}
\end{figure}

    Unfortunately, the intrinsic distribution of host extinctions of CCSNe in LIRGs is uncertain. Using radio observations \cite{MM01} estimated extinctions towards 19 SN remnants in the central 500 pc region of the nearby IR bright starburst galaxy M~82 finding a mean extinction of $A_{V} = 24 \pm9$ mag. So far, the largest extinction towards a SN discovered at IR wavelengths was reported by \cite{iras17138_SN} for SN 2008cs discovered at 1.5 kpc projected distance from the centre of the LIRG IRAS 17138-1017.
    However, the exact shape of the extinction distribution becomes irrelevant for our study at very high extinction, for example any SN suffering from a host extinction of $A_{V} \gtrsim 40$ mag is equally unobservable in our data set. Thus the shape of the high host extinction distribution is not important, but rather the fraction of the high extinction events.
    The host extinction of the simulated SNe in the Monte Carlo simulation was split into two components: A "surface component" fraction taken into account with the extinction model, and an extremely obscured "deep component" that corresponds to the missing fraction. For the surface component, we fitted a set of host galaxy extinctions for 19 CCSNe discovered by previous optical or near-IR searches in central regions of LIRGs from \cite{host_extinction_paper} with an exponential function. The best fit resulted in a function $P(A_V)=0.881e^{-0.084A_V}$, with a median probability of $A_{V} = 7.7$ mag (Fig. \ref{extinction_figure}). The extinction values applied to the simulated CCSNe were randomly selected from this distribution. SNe in the missing fraction are considered to be completely undetectable, exploding at extremely dusty nuclear and circum-nuclear regions of their host galaxy. The probability was set before each run for a simulated SN to belong either to the surface extinction fraction or to the missing fraction, with the aim to constrain the latter. This missing fraction probability was varied between 0\% and 99\%. Galactic extinctions were based on the dust map calibration of \cite{schlafly2011} for the sample galaxies and were acquired from NASA/IPAC Extragalactic Database (NED), and are listed in Table \ref{data set_table}.
    For comparison, the SED models used to derive the intrinsic CCSN rates in Sect. \ref{intrinsic_CCSN_rates} assume the following extinction ranges for the different galaxy components: $A_V \lesssim 20$ mag for the diffuse dust component, $A_V \sim 50 - 250$ mag for the star-forming regions and giant molecular clouds, and $A_V \sim 250 - 1500$ mag for the AGN torus.

    \subsubsection{Control times}
    \label{control_times}

    Total control time (TCT) is a measure of how long a galaxy has had an opportunity to produce detectable SNe within a survey \citep{control_time_original,cappellaro1993}, and is computed from individual control times (CT). For example, a five year survey with one year cadence would not automatically result in a TCT of five years, since many SNe could have potentially exploded and faded between the observations. Calculation of the CTs requires information on the absolute peak magnitude distribution and the light curve evolution for a given SN type, distance modulus, line-of-sight extinction, and the limiting magnitudes of the image. Therefore, the CT describes how long on average a SN could be detected. The descriptions of the adopted SN absolute peak magnitude distributions and light curve evolution were provided in the previous section. 
    
    For limiting magnitudes, each image epoch was handled separately. Each pixel in the limiting magnitude image was multiplied by a weight, equal to the ratio of the flux of the pixel compared to the total flux of the galaxy, and then their mean was calculated; this mean value was adopted as the limiting magnitude of the epoch for the purpose of the CT calculation. Subsequently, a Monte Carlo method was used to generate a total of $10^5$ light curves of each SN subtype by applying absolute peak magnitude, time-scale randomization and random host extinction on the template light curves based on the parameters described in Table \ref{sn_models}. For each light curve it was measured for how long the apparent magnitude of the SN was brighter than the flux weighted average limiting magnitude for that epoch. Finally, the CT for that epoch was estimated as the mean of those times. This process was repeated for each galaxy and each epoch in the data set. Variance in repeated runs of this process was less than one day, which was the time resolution of the main Monte Carlo method.
    
    To calculate the TCTs, the time differences between the observed epochs were summed, except if the time difference was larger than the CT, in which case the CT was used instead of the time difference. The CTs and TCTs were calculated for each CCSN subtype separately. The start date for the detection probability Monte Carlo simulation was the beginning date of the CT of the first observed epoch. In the process CCSNe were simulated only within the CT ranges. The median CT of the fastest evolving CCSN template IIb was 71 days. 

    \subsubsection{Intrinsic CCSN rates}
    \label{intrinsic_CCSN_rates}

    Intrinsic CCSN rates for the data set were obtained with an adapted version of SED Analysis Through Markov Chains (SATMC) Monte Carlo code \citep{SATMC,efstathiou2022}. The LIRG SEDs were constructed from a combination of archival photometry and spectra. Included were data from the Galaxy Evolution Explorer (GALEX), 2MASS, the Infrared Astronomical Satellite (IRAS), and the Infrared Space Observatory (ISO) photometric data via the NASA/IPAC Extragalactic Database (NED) and references therein. Photometric data was also obtained from the Data Release 2 of the Panoramic Survey Telescope and Rapid Response System (Pan-STARRS; \citealt{chambers2016}), and the Herschel Space Observatory image atlas by \cite{chu2017}. Furthermore, Spitzer spectra used in a SED analysis by \cite{herrero2017} were also employed. 
    
\begin{figure*}
\centering
\includegraphics[trim={0 0 0 0.2cm},clip,width=0.49\linewidth]{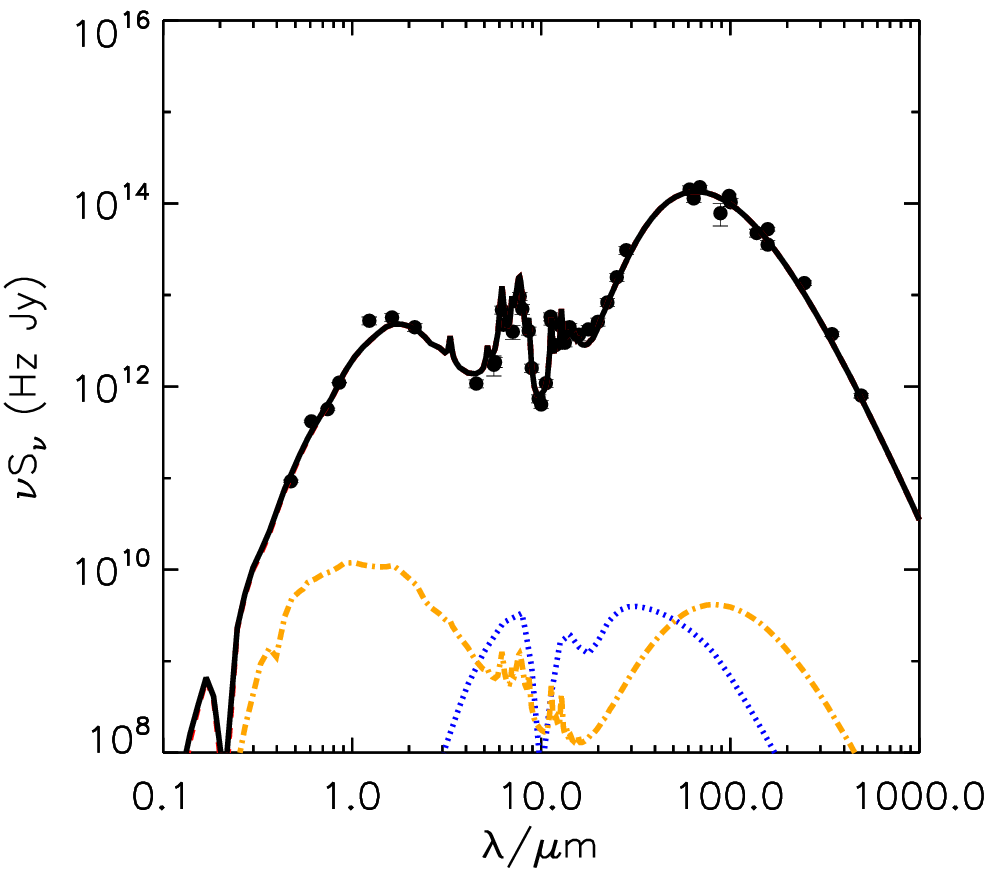}
\includegraphics[trim={0 0 0 0.1cm},clip,width=0.49\linewidth]{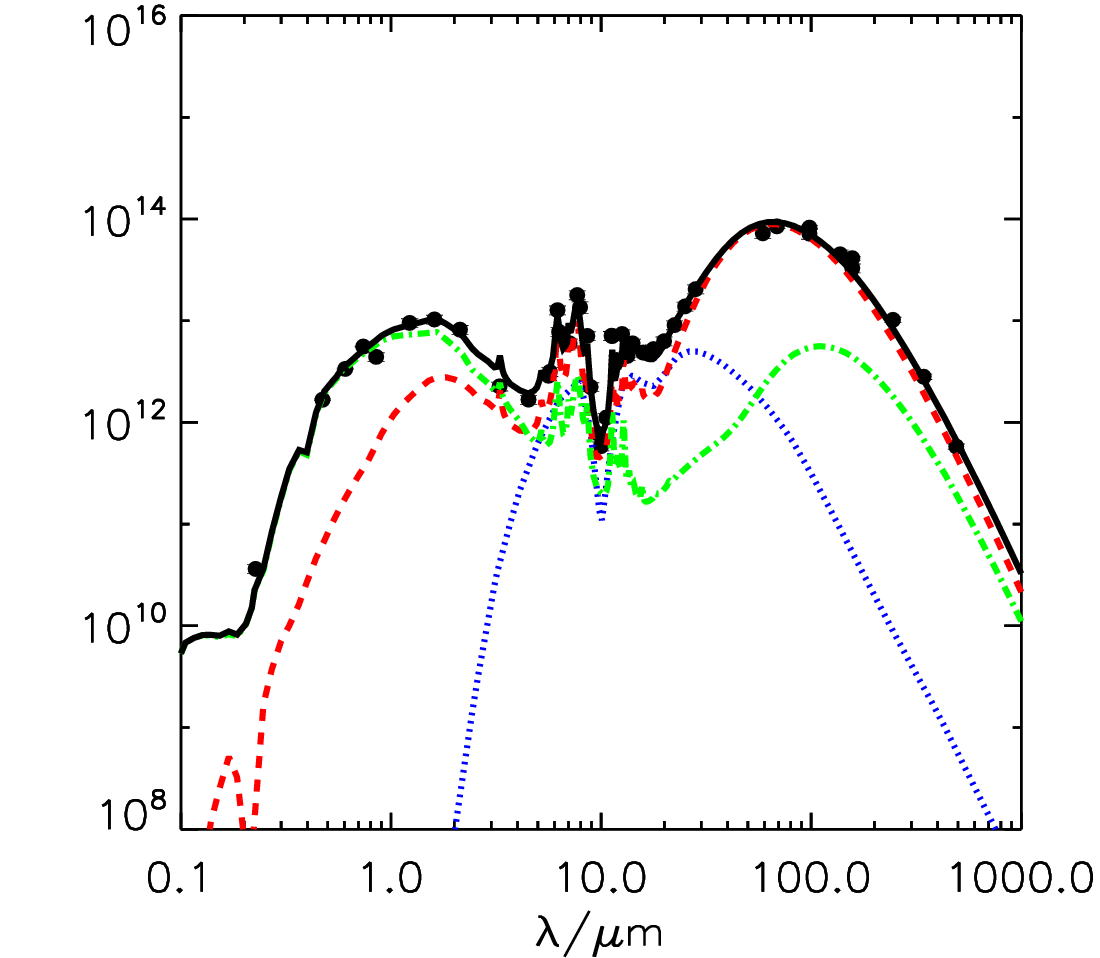}
\includegraphics[trim={0 0 0 0.15cm},clip,width=0.49\linewidth]{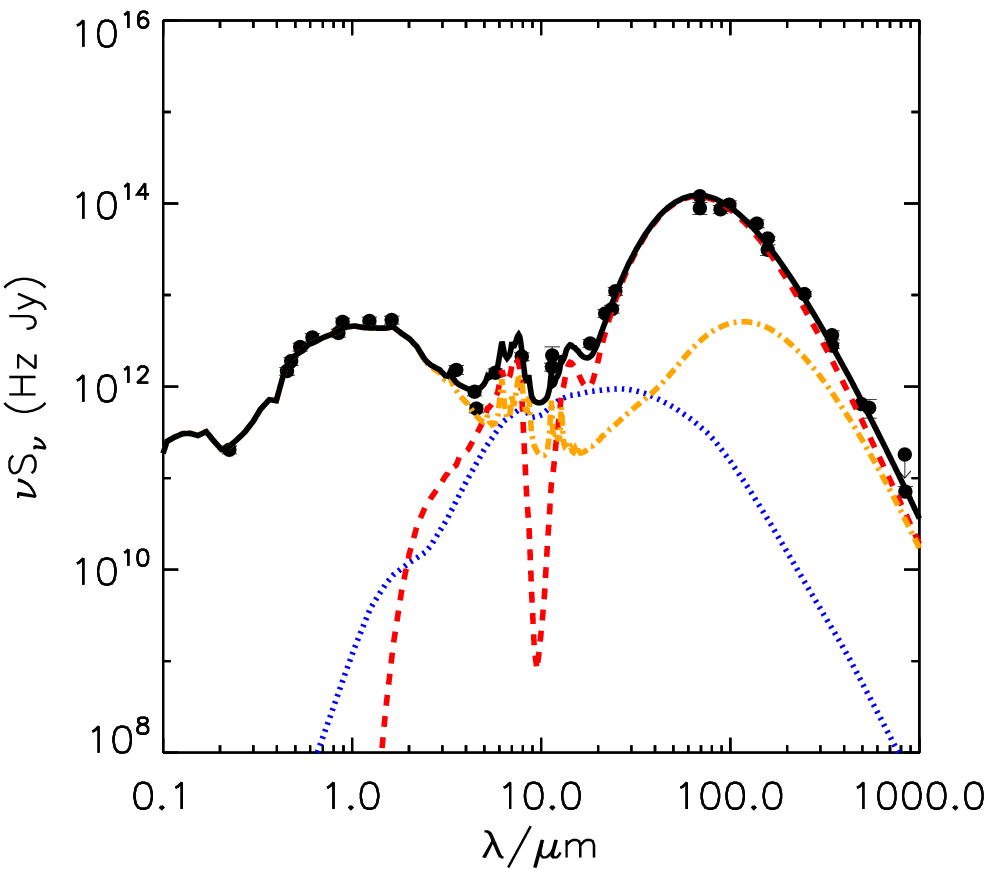}
\includegraphics[trim={0 0 0 0.1cm},clip,width=0.49\linewidth]{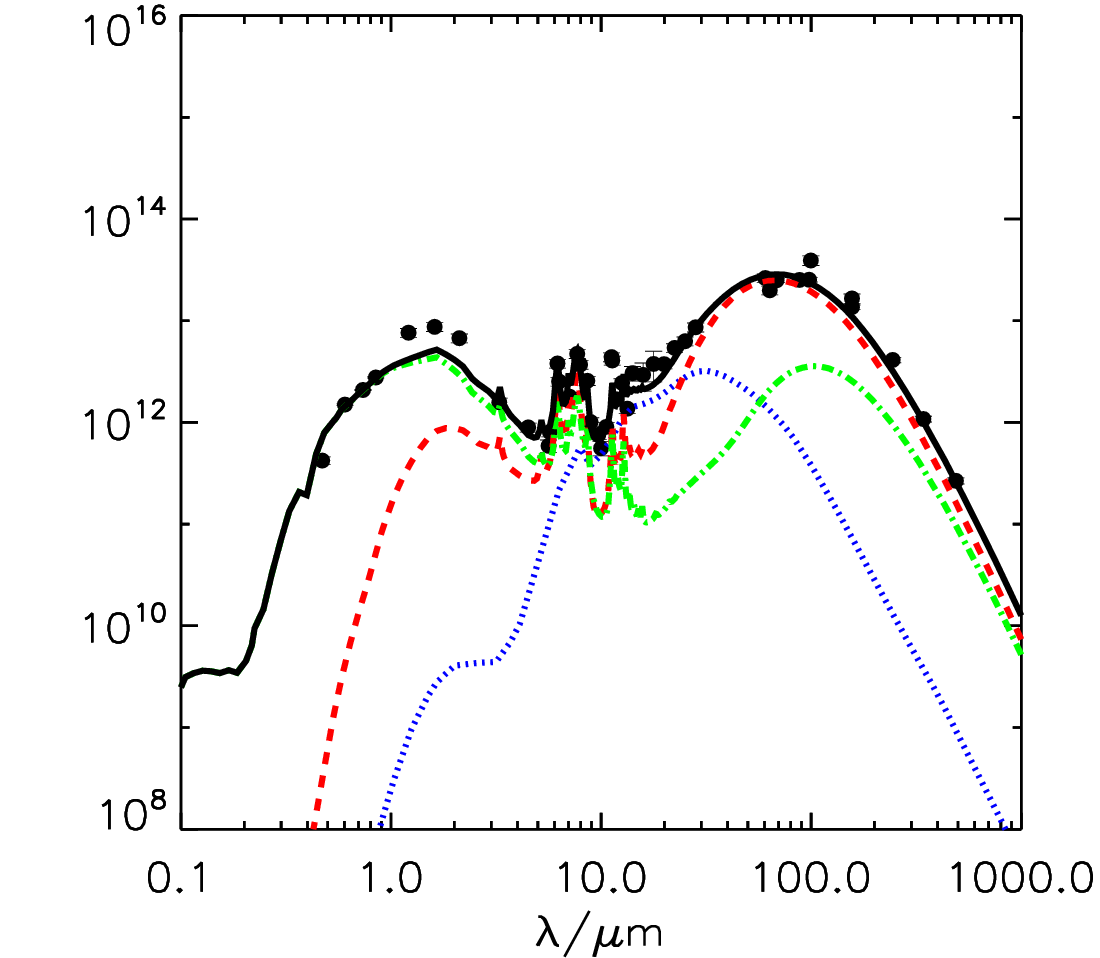}
\caption{Our SED models (solid black curves) for the observations (black points) of four sample LIRGs, including IRAS 17578-0400 (top left), MCG+08-11-002 (top right), CGCG 049-057 (bottom left), and IRAS 16516-0948 (bottom right). The model components include a spheroidal (dot-dashed orange) or a disc (dot-dashed green) galaxy, starburst contribution (dashed red), and an AGN (dotted blue). We note that the starburst component of IRAS 17578-0400 dominates the total emission and is overlapped by the combined model curve, due to the small contribution of the other components.}
\label{sed_examples}
\end{figure*}

    The components of the radiative transfer model were as follows: starburst \citep{SED_starburst1,SED_starburst2}, AGN \citep{SED_AGN1,SED_AGN2} and a spheroidal galaxy \citep{SED_spheroid} or a disc galaxy \citep{efstathiou2022}. SATMC method fits simultaneously the different model components to the SED of a galaxy to estimate the contribution of each component. The contribution of the starburst component constrains the total mass and age of the starburst. The model adopts the Salpeter initial mass function (IMF), solar metallicity and stellar population synthesis models of \cite{bruzual2003}. The model determines the starburst age from the SED fit and then follows the stellar population models to that age to determine the CCSN rate. The model assumes that the massive stars on their stellar tracks can explode as CCSNe up to an age of $\sim$42.6 Myr, corresponding to masses of $\sim8 M_{\odot}$. However, since the starburst ages derived for the LIRGs in the data set are younger than this, the exact lower mass limit of CCSNe is not critical. The intrinsic CCSN rate estimates were derived in this work for CGCG 049-057, IRAS 16516-0948, IRAS 17578-0400 and MCG+08-11-002 with the SED fits shown in Fig. \ref{sed_examples}. Intrinsic CCSN rates for IRAS 17138-1017, IC 883 and Arp 299-A/B were obtained from \cite{host_extinction_paper} and derived with the same method. Intrinsic CCSN rate for Arp 299-A/B was obtained for the combined system, which was divided between the A and B systems based on results by \cite{mattila2012}. The resulting CCSN rates, SFRs and starburst ages from the SATMC are listed in Table \ref{data set_table}.

\begin{figure}[hbt]
    \begin{center}
        \includegraphics[width=0.5\textwidth]{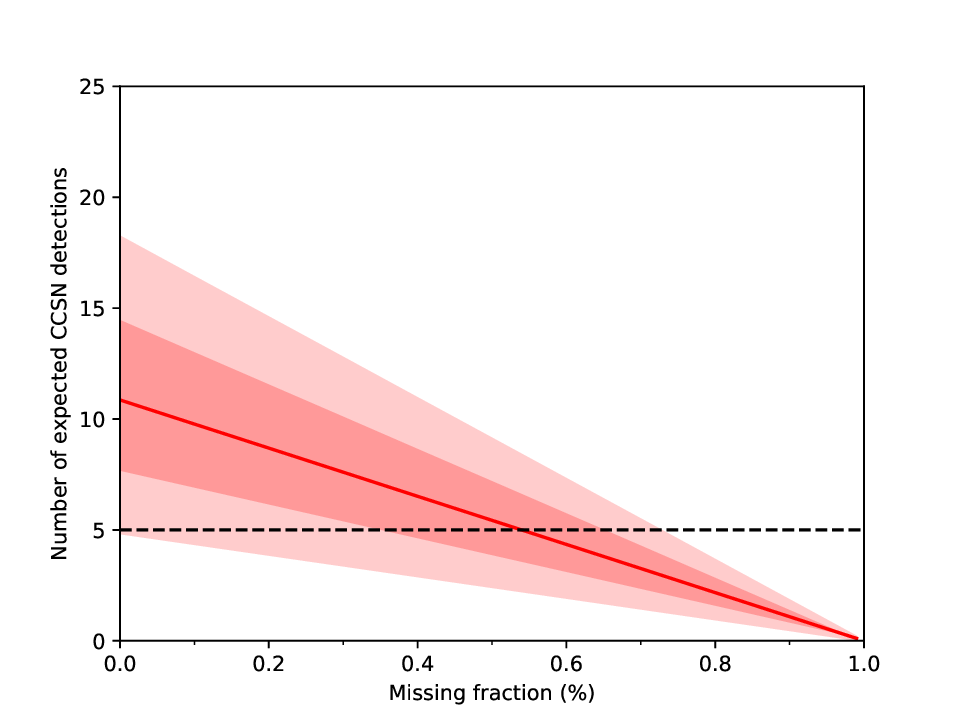}
        \end{center}
    \caption{Number of expected CCSNe detections from the Monte Carlo simulation compared to the missing fraction. Red line: Total number of expected CCSN detections from the data set over the survey period as a function of the missing fraction. Shaded red areas: 1$\sigma$ and 2$\sigma$ confidence areas for the total number of expected CCSN detections from the data set over the survey period as a function of the missing fraction. Poissonian upper and lower limits due to small number statistics and the cumulative effect of other error sources are combined here. Dashed black: Number of real detected CCSNe in the survey.}
    \label{missing_fraction_figure}
\end{figure}

\subsection{Error estimation}
\label{error_estimation}

    Due to the random nature of the Monte Carlo method, the impact of the individual sources of error become chaotic. The effect of this was estimated by running the main Monte Carlo program with selected parameters offset by their respective errors, for example the CCSN template peak luminosities increased by their 1$\sigma$ variance, to evaluate its effect on the results. The relevant parameters and their respective errors are the following. First, the process of determining the limiting magnitudes for the images used a step of 0.2 mag, giving an error of $\pm 0.2$ mag. Next, for the photometric zero point calibration a conservative error of $\pm$0.3 mag was estimated based on the largest difference between the photometric zero point corrections across the data set. Another uncertainty to take into account is the error of the $M_R - M_K$ colour difference to estimate the \textit{K}-band peak magnitudes based on the \textit{R}-band absolute peak magnitudes. H-poor templates had a standard deviation of $\pm 0.26$ mag and Type II templates a standard deviation of $\pm 0.07$ mag. Since the 87A-like and IIn templates had only one light curve on which the shift was based, they had no statistical error to use. Conservatively, we adopted the largest errors for this uncertainty. Finally, the error on the distance modulus for the galaxies from NED was $\pm 0.15$ mag. Thus, the overall error as the square root of the sum of squares is $\pm 0.47$ mag. 

    The number of intrinsic CCSNe predicted by the Monte Carlo simulation is small, and thus sensitive to small number statistics. Therefore, we adopt Poissonian confidence limits by \cite{gehrels1986} to the estimated number of events as their uncertainty.

\section{Results}
\label{results}

    The output of the Monte Carlo simulation is an average detection probability of SNe for each galaxy (i.e. the fraction of detectable CCSNe from the estimated intrinsic number of CCSNe, as a function of the missing fraction). These values are presented for all the sample galaxies in Table \ref{detection_prob_table_high} for a selection of missing fraction values. The SATMC method has provided us with intrinsic CCSN rates for the galaxies in our sample. The intrinsic CCSN rates are split amongst all the CCSN subtypes according to rates by \cite{SN_subtype_rates}, and then multiplied by their respective TCTs to acquire the intrinsic number of CCSNe of each subtype. An example for IC 883: The total CTs of different CCSN subtypes are 1.70, 1.82, 1.95, 2.16, 2.48 and 4.77 yrs for the Type IIb, Ib, Ic, II, 87A-like and IIn SNe, respectively. The intrinsic CCSN rate is 2.1 SN yr$^{-1}$, which corresponds to 0.30, 0.34, 0.32, 2.79, 0.17 and 1.0 SN after adjusting the relative rates for the difference in CTs, and multiplying by the corresponding CTs. However, only a fraction of these are detected due to the combined effects of host extinction and light curve evolution bringing the SN brightness below the limiting magnitude at the explosion position. The resulting values are presented in Table \ref{possible_sn_table_high} for a selection of missing fraction values. Figure \ref{missing_fraction_figure} plots expected CCSN detections as a function of the missing fraction. 
    
    Based on the number of five detected CCSNe in the survey, we obtain a missing fraction of $f_{\mathrm{mis}} = 54.0^{+11.6}_{-19.7}(^{+18.5}_{-54.0})$\% with 1$\sigma$ and (2$\sigma$) uncertainties. This is the missing fraction of CCSNe which have extremely high host extinctions and thus cannot be discovered by optical or NIR observations. Additionally, CCSNe not included in the missing fraction may also remain undiscovered due to moderate host extinctions. The fraction of these moderately obscured and undiscovered CCSNe depends on the survey depth and can be calculated by integrating the host extinction model curve over the relevant extinction range. Within the compiled sample of CCSNe discovered in the central regions of LIRGs by \cite{host_extinction_paper}, the largest estimated host galaxy extinction among the reported optically discovered CCSNe was $A_V = 3.1$ mag. Therefore, as an example, in the context of a survey that is able to detect objects up to $A_V = 3$ mag, the fraction of CCSNe with $A_V \leq 3$ mag and $A_V > 3$ mag is 22 \% and 78 \%, respectively, within the CCSNe not included in the missing fraction, resulting in the total fraction of undetectable CCSNe of $89.7^{+2.6}_{-4.4}$ \%. Similarly for near-infrared surveys, choosing extinction threshold examples of $A_V = 13$ and $A_V = 16$ mag, we obtain the total fractions of undetectable CCSNe of $69.4^{+7.7}_{-13.1}$ and $66.0^{+8.6}_{-14.6}$ \%, respectively. The limit $A_V = 16$ was chosen as the highest host extinction of a SN in the data set. The limit $A_V = 13$ mag was calculated to be representative of the ALTAIR/NIRI data set on average. The median limiting magnitude of the data set was 19.4 mag, and the median luminosity distance was 74.7 Mpc. The extinction limit was then chosen as the amount of host extinction after which the most common CCSN subtype template, Type II, would remain detectable for its median CT of 100 days. The median galactic extinction $A^{\mathrm{Gal}}_K$ for the data set was 0.05 mag, and was thus assumed to be zero in this calculation.

    In Table \ref{host_ext_missing_fracion} we list the detected fraction of CCSNe with host extinctions within the $P(A_V)$ distribution below a detection threshold $A_{V,\mathrm{limit}}$ calculated as $f_{A_V} = \int_{0}^{A_{V,limit}} P(A_V)\,dA_V$; the fraction of CCSNe that remain undetected due to host extinction described by the host extinction model calculated as $f_{\mathrm{ext}} = [(1 - f_{A_V}) \times (1 - f_{\mathrm{mis}})]$; and the total undetected fraction calculated as the sum of the missing fraction and the fraction of CCSNe that remain undetected due to host extinction described by the host extinction model, $f_{\mathrm{tot}} = f_{\mathrm{ext}} + f_{\mathrm{mis}}$. Surveys with deeper detection thresholds are able to detect more of the CCSN population not included in the missing fraction, which is assumed to remain always undetected.

    We additionally note that the median CT for the fastest evolving CCSN template (Type IIb) is 71 days. This suggests that a minimum cadence of $\sim$2 months is optimal for a similar survey as our data set to expect to detect half of the Type IIb SNe that could be discovered in at least one epoch, assuming minimal extinction in \textit{K}-band.

    Comparing our undetectable fraction estimate of $66.0^{+8.6}_{-14.6}$ \% for infrared surveys with a limit of $A_V = 16$ mag, we find very similar undetectable fraction to the one by \cite{miluzio2013} of $\sim 60\%$ to $75\%$. This is not surprising since we have very similar methods. Their galaxy sample is also similar including 30 starburst galaxies of which only one is a non-(U)LIRG and two are ULIRGs. They also performed their analysis in \textit{K}-band, with a notable difference of using non-AO data for this. 

    \cite{mattila2012} estimated that $83^{+9}_{-15}$ \% of CCSNe remain undetected by optical observations in LIRGs due to high host extinctions based on the reported discoveries of CCSNe in the nearby LIRG Arp 299. Assuming an optical survey extinction limit of 3 mag, our total fraction of undetectable CCSNe results in $89.7^{+2.6}_{-4.4}$\%, which is in good agreement with the results of \cite{mattila2012}. The intrinsic CCSN rate of Arp 299 in their work was calculated with SED fitting similarly to our work, but the A and B cores were treated separately, and the SED data was limited in wavelength coverage. The CCSN rate of the C nucleus (in combination with the C' component) and the circumnuclear regions were determined from the empirical equation by \cite{MM01} which estimates the CCSN rate based on the infrared luminosity of the LIRG. This resulted in a CCSN rate of 1.29 and 0.30-0.61 CCSN $yr^{-1}$ for the nuclei and the circumnuclear regions, respectively. This work adopts the more recent results from \cite{host_extinction_paper} where the SED fit was carried out for the whole Arp 299 system with one SED that also includes a broader wavelength range, which resulted in a total CCSN rate of $1.5^{+0.1}_{-0.2}$ CCSN yr$^{-1}$. The resulting CCSN rate was divided in this work between the A and B+C nuclei based on their IR luminosities.

    Based on their results, \cite{fox2021} missed $\sim$83\% of the CCSNe in their sensitivity range. They estimated the intrinsic CCSN rates based on infrared luminosity with the equation by \cite{MM01}. Their undetected fraction of $\sim$83\% is significantly higher than our 66\% that we derive for $A_V = 16$ mag host extinction limit, despite the similar wavelength of the search. They explained that the poor resolution of Spitzer and its asymmetric PSF resulted in strong image subtraction residuals and low sensitivity at the central regions of the sample galaxies. Therefore, a comparably higher undetected fraction can be expected.

    The extinction calculations in this work have assumed the commonly used \cite{cardelli1989} extinction law with $R_V = 3.1$, which results in $A_K/A_V = 0.114$. For comparison, we calculated some of the survey thresholds with alternative reddening laws, including the \cite{cardelli1989} extinction law with $R_V$ = 2.7 (Small Magellanic Cloud) and $R_V$ = 4.05 (starforming galaxy), the \cite{calzetti2000} attenuation law with $R_V$ = 4.05, and the \cite{fitzpatrick1999} extinction law with $R_V$ = 3.1. These resulted in $A_K/A_V$ ratios of 0.107, 0.124, 0.091 and 0.116, respectively. Converting these ratios for $A_V$ = 16 mag results in $A_K$ values of 1.8 mag for the \cite{cardelli1989} law with $R_V$ = 3.1, and 1.7, 2.0, 1.5, 1.9 mag for the rest, respectively. Therefore, the differences are not large and are basically within one 0.2 mag step in our detection threshold simulations.

\begin{table*}[h]
\begin{center}
\caption{The estimated percentage of recoverable CCSNe in the survey data of the sample galaxies, for different adopted missing fraction values.}
\begin{tabular}{cccccccccccc}
\hline
\hline
missing fraction & 0\% & 10\% & 20\% & 30\% & 40\% & 50\% & 60\% & 70\% & 80\% & 90\% & 99\% \\
\hline
Arp 299-A & 0.80 & 0.72 & 0.64 & 0.56 & 0.48 & 0.40 & 0.32 & 0.24 & 0.16 & 0.08 & 0.01 \\
Arp 299-B & 0.45 & 0.40 & 0.36 & 0.31 & 0.27 & 0.22 & 0.18 & 0.13 & 0.09 & 0.04 & 0.00 \\
CGCG 049-057 & 0.82 & 0.74 & 0.66 & 0.57 & 0.49 & 0.41 & 0.33 & 0.25 & 0.16 & 0.08 & 0.01 \\
IC 883 & 0.55 & 0.49 & 0.44 & 0.38 & 0.33 & 0.27 & 0.22 & 0.16 & 0.11 & 0.05 & 0.01 \\
IRAS 16516-0948 & 0.59 & 0.53 & 0.47 & 0.41 & 0.35 & 0.29 & 0.24 & 0.18 & 0.12 & 0.06 & 0.01 \\
IRAS 17138-1017 & 0.63 & 0.57 & 0.51 & 0.44 & 0.38 & 0.32 & 0.25 & 0.19 & 0.13 & 0.06 & 0.01 \\
IRAS 17578-0400 & 0.72 & 0.64 & 0.57 & 0.50 & 0.43 & 0.36 & 0.29 & 0.21 & 0.14 & 0.07 & 0.01 \\
MCG+08-11-002 & 0.69 & 0.62 & 0.55 & 0.48 & 0.42 & 0.35 & 0.28 & 0.21 & 0.14 & 0.07 & 0.01 \\
\hline
\end{tabular}
\label{detection_prob_table_high}
\end{center}
\end{table*}

\begin{table*}[h]
\begin{center}
\caption{Expected number of detected CCSNe for each galaxy in the data set, for different missing fractions.}
\begin{tabular}{cccccccccccc}
\hline
\hline
missing fraction & 0\% & 10\% & 20\% & 30\% & 40\% & 50\% & 60\% & 70\% & 80\% & 90\% & 99\% \\
\hline
Arp 299-A & 2.22 & 2.00 & 1.78 & 1.56 & 1.33 & 1.11 & 0.89 & 0.67 & 0.44 & 0.22 & 0.02 \\
Arp 299-B & 0.63 & 0.57 & 0.50 & 0.44 & 0.38 & 0.32 & 0.25 & 0.19 & 0.13 & 0.06 & 0.01 \\
CGCG 049-057 & 1.38 & 1.24 & 1.11 & 0.97 & 0.83 & 0.69 & 0.55 & 0.41 & 0.28 & 0.14 & 0.01 \\
IC 883 & 2.70 & 2.43 & 2.16 & 1.89 & 1.62 & 1.35 & 1.08 & 0.81 & 0.54 & 0.27 & 0.03 \\
IRAS 16516-0948 & 0.44 & 0.39 & 0.35 & 0.31 & 0.26 & 0.22 & 0.18 & 0.13 & 0.09 & 0.04 & 0.00 \\
IRAS 17138-1017 & 0.88 & 0.79 & 0.71 & 0.62 & 0.53 & 0.44 & 0.35 & 0.26 & 0.18 & 0.09 & 0.01 \\
IRAS 17578-0400 & 1.66 & 1.50 & 1.33 & 1.16 & 1.00 & 0.83 & 0.66 & 0.50 & 0.33 & 0.17 & 0.02 \\
MCG+08-11-002 & 0.94 & 0.85 & 0.76 & 0.66 & 0.57 & 0.47 & 0.38 & 0.28 & 0.19 & 0.09 & 0.01 \\
Total & 10.86& 9.77 & 8.69 & 7.60 & 6.52 & 5.43 & 4.35 & 3.26 & 2.17  &1.09 & 	0.11 \\

\hline
\end{tabular}
\label{possible_sn_table_high}
\end{center}
\end{table*}

\begin{table}[h]
\begin{center}
\caption{The fraction of CCSNe with host extinctions below the extinction threshold, $f_{A_V}$, the fraction of CCSNe that remain undetected due to host extinction described by the host extinction model, $f_{\mathrm{ext}}$, and the total undetectable fraction, $f_{\mathrm{tot}}$, assuming a missing fraction of $f_{\mathrm{mis}} = 54.0^{+11.6}_{-19.7}$ \% and survey parameters similar to our data set.}
\begin{tabular}{cccc}
\hline
\hline
$A_{V,\mathrm{limit}}$ & $f_{A_V}$ & $f_{\mathrm{ext}}$ & $f_{\mathrm{tot}}$\\
(mag) & (\%) & (\%) & (\%)\\
\hline
2 & 15.5 & $38.9$ & $92.9^{+1.8}_{-3.1}$\\
3& 22.3 & $35.7$ & $89.7^{+2.6}_{-4.4}$ \\
4& 28.5 & $32.9$ & $86.9^{+3.3}_{-5.6}$ \\
5& 34.3 & $30.2$ & $84.2^{+4.0}_{-6.8}$ \\
7& 44.5 & $25.5$ & $79.5^{+5.2}_{-8.8}$ \\
10& 56.8 & $19.8$ & $73.8^{+6.6}_{-11.2}$ \\
13& 66.4 & $15.4$ & $69.4^{+7.7}_{-13.1}$ \\
16& 73.9 & $12.0$ & $66.0^{+8.6}_{-14.6}$ \\
20& 81.4 & $8.6$ & $62.6^{+9.4}_{-16.0}$ \\
25& 87.7 & $5.6$ & $59.6^{+10.2}_{-17.3}$ \\
30& 92.0 & $3.7$ & $57.7^{+10.7}_{-18.1}$ \\
40& 96.4 & $1.6$ & $55.6^{+11.2}_{-19.0}$ \\
\hline
\end{tabular}
\label{host_ext_missing_fracion}
\end{center}
\end{table}

\section{Conclusions}
\label{conclusions}

    We have presented new estimates for the fraction of undetectable CCSNe in LIRGs in the local Universe with a detailed description of our method. The analysis is based on the SUNBIRD survey near-IR $K$-band monitoring data set of eight LIRGs using the Gemini-North Telescope with the ALTAIR/NIRI laser guide star AO system, which has yielded detections of five CCSNe. We have also presented a new extinction distribution for CCSNe in the central regions of LIRGs $P(A_V)=0.881e^{-0.084A_V}$ (with a median probability of $A_{V} = 7.7$), based on the distribution of estimated extinctions for CCSNe discovered in the central regions of LIRGs at optical and near-IR wavelengths \citep{host_extinction_paper}. This distribution still excludes those CCSNe that have the highest extinctions and are accounted for in the missing SN fraction. Based on the analysed SUNBIRD data set, the resulting missing fraction for CCSNe in LIRGs is $54.0^{+11.6}_{-19.7}$\%. 
    
    The total fraction of undetectable CCSNe depends on the survey detection threshold for the host galaxy extinction. The additional contribution to the undetectable fraction can be calculated by integrating over our adopted extinction model probability distribution above the extinction limit. For assumed typical examples of host extinction limits for near-infrared ($A_V$ = 16 mag) and optical ($A_V$ = 3 mag) surveys result in a total undetectable fraction of $66.0^{+8.6}_{-14.6}$\% and $89.7^{+2.6}_{-4.4}$\%, respectively.

    Transient surveys locally and at intermediate redshifts will be revolutionised by next-generation programmes like the Legacy Survey of Space and Time (LSST) of the Vera C. Rubin Observatory whereas the JWST will be able to probe the CCSN rates in the cosmic noon. Estimates for the undetected fraction of CCSNe in local LIRGs will be crucial for calibrating the CCSN rates derived by the LSST, JWST and other next-generation facilities.
    
\begin{acknowledgements}
    We thank the anonymous referee for useful comments.

    I.M. and E.K. acknowledge financial support from the Emil Aaltonen foundation.

    T.M.R. is part of the Cosmic Dawn Center (DAWN), which is funded by the Danish National Research Foundation under grant DNRF140. S.M. and T.M.R. acknowledge support from the Research Council of Finland project 350458. 

    C.V. acknowledges financial support from the Vilho, Yrjö and Kalle Väisälä Fund.
    
    Based on observations obtained at the international Gemini Observatory, a program of NSF NOIRLab, which is managed by the Association of Universities for Research in Astronomy (AURA) under a cooperative agreement with the U.S. National Science Foundation on behalf of the Gemini Observatory partnership: the U.S. National Science Foundation (United States), National Research Council (Canada), Agencia Nacional de Investigaci\'{o}n y Desarrollo (Chile), Ministerio de Ciencia, Tecnolog\'{i}a e Innovaci\'{o}n (Argentina), Minist\'{e}rio da Ci\^{e}ncia, Tecnologia, Inova\c{c}\~{o}es e Comunica\c{c}\~{o}es (Brazil), and Korea Astronomy and Space Science Institute (Republic of Korea).

    Based on observations made with the Nordic Optical Telescope, owned in collaboration by the University of Turku and Aarhus University, and operated jointly by Aarhus University, the University of Turku and the University of Oslo, representing Denmark, Finland and Norway, the University of Iceland and Stockholm University at the Observatorio del Roque de los Muchachos, La Palma, Spain, of the Instituto de Astrofisica de Canarias.

    This publication makes use of data products from the Two Micron All Sky Survey, which is a joint project of the University of Massachusetts and the Infrared Processing and Analysis Center/California Institute of Technology, funded by the National Aeronautics and Space Administration and the National Science Foundation.

    This research has made use of the NASA/IPAC Extragalactic Database (NED), which is operated by the Jet Propulsion Laboratory, California Institute of Technology, under contract with the National Aeronautics and Space Administration.
\end{acknowledgements}

\bibliography{sources}

\begin{appendix}

\onecolumn
\section{Additional tables and figures}

\begin{table*}[h]
\begin{center}
\caption{Survey epochs and CTs of Arp 299-A.}
\begin{tabular}{ccccccccc}
\hline
\hline
    MJD & $m_{\mathrm{Lim}}$  & $\Delta t$  & IIb & Ib & Ic& II & 87A & IIn\\
    &(mag)& (d)  &  (d)& (d)& (d)& (d)& (d) & (d) \\
\hline
54637.3 & 19.26 & - & 137 & 146 & 154 & 195 & 245 & 910 \\
54879.5 & 19.51 & 242 & 149 & 157 & 165 & 212 & 260 & 981 \\
54958.3 & 19.50 & 79 & 148 & 157 & 164 & 211 & 259 & 976 \\
55168.6 & 19.15 & 210 & 133 & 141 & 150 & 189 & 235 & 873 \\
55223.6 & 17.64 & 55 & 72 & 80 & 87 & 101 & 131 & 448 \\
55251.4 & 18.77 & 28 & 118 & 125 & 134 & 165 & 206 & 771 \\
55321.3 & 19.51 & 70 & 148 & 157 & 165 & 212 & 261 & 975 \\
55955.5 & 19.84 & 634 & 163 & 171 & 179 & 233 & 284 & 1075 \\
\hline
\end{tabular}
\label{CT_ic694_high}
\end{center}
\end{table*}

\begin{table*}[h]
\begin{center}
\caption{Survey epochs and CTs of Arp 299-B.}
\begin{tabular}{ccccccccc}
\hline
\hline
MJD & $m_{\mathrm{Lim}}$ & $\Delta t$ & IIb & Ib & Ic & II & 87A & IIn \\
    &(mag)& (d)  &  (d)& (d)& (d)& (d)& (d) & (d) \\
\hline
54165.4 & 18.25 & - & 84 & 92 & 100 & 118 & 150 & 532 \\
54639.3 & 18.06 & 474 & 75 & 85 & 92 & 107 & 138 & 478 \\
54817.6 & 17.39 & 178 & 52 & 59 & 65 & 76 & 98 & 301 \\
54928.3 & 17.23 & 111 & 46 & 52 & 59 & 68 & 87 & 264 \\
54958.3 & 17.59 & 30 & 59 & 66 & 73 & 84 & 111 & 350 \\
55170.6 & 16.04 & 212 & 15 & 14 & 20 & 19 & 31 & 68 \\
55223.6 & 17.05 & 53 & 41 & 45 & 53 & 59 & 76 & 223 \\
55252.4 & 16.76 & 29 & 32 & 35 & 42 & 48 & 61 & 167 \\
55317.3 & 16.85 & 65 & 35 & 38 & 45 & 51 & 65 & 185 \\
55668.3 & 18.25 & 351 & 84 & 92 & 99 & 118 & 152 & 529 \\
55955.6 & 16.64 & 287 & 29 & 31 & 38 & 43 & 55 & 147 \\
\hline
\end{tabular}
\label{CT_arp299_high}
\end{center}
\end{table*}

\begin{table*}[h]
\begin{center}
\caption{Survey epochs and CTs of CGCG 049-057 for each CCSN subtype.}
\begin{tabular}{ccccccccc}
\hline
\hline
    MJD & $m_{\mathrm{Lim}}$ & $\Delta t$ & IIb & Ib & Ic & II & 87A & IIn \\
    &(mag)& (d)  &  (d)& (d)& (d)& (d)& (d) & (d) \\
\hline
54578.5 & 19.30 & - & 101 & 110 & 117 & 141 & 181 & 647 \\
54637.3 & 19.46 & 59 & 107 & 116 & 124 & 150 & 190 & 686 \\
54879.6 & 20.13 & 242 & 135 & 144 & 152 & 191 & 239 & 883 \\
54956.5 & 20.03 & 77 & 131 & 141 & 147 & 186 & 231 & 859 \\
54989.4 & 20.05 & 33 & 131 & 142 & 148 & 186 & 232 & 854 \\
55251.6 & 19.96 & 262 & 128 & 136 & 146 & 181 & 226 & 831 \\
55308.4 & 20.05 & 57 & 131 & 140 & 149 & 186 & 230 & 858 \\
56145.3 & 20.03 & 837 & 131 & 139 & 148 & 184 & 230 & 847 \\
\hline
\end{tabular}
\label{CT_cgcg_high}
\end{center}
\end{table*}

\begin{table*}[h]
\begin{center}
\caption{Survey epochs and CTs of IC 883.}
\begin{tabular}{ccccccccc}
\hline
\hline
MJD & $m_{\mathrm{Lim}}$ & $\Delta t$ & IIb & Ib & Ic & II & 87A & IIn \\
    &(mag)& (d)  &  (d)& (d)& (d)& (d)& (d) & (d) \\
\hline
54571.5 & 19.26 & - & 59 & 67 & 73 & 86 & 110 & 356 \\
54637.3 & 19.48 & 66 & 67 & 74 & 82 & 96 & 124 & 405 \\
54861.6 & 19.00 & 224 & 51 & 56 & 64 & 74 & 94 & 292 \\
54956.4 & 19.45 & 95 & 66 & 74 & 81 & 95 & 120 & 398 \\
54992.4 & 19.48 & 36 & 68 & 75 & 82 & 96 & 122 & 417 \\
55251.5 & 19.41 & 259 & 65 & 71 & 80 & 92 & 117 & 389 \\
55320.5 & 19.26 & 69 & 60 & 65 & 74 & 87 & 109 & 351 \\
55351.3 & 19.45 & 31 & 66 & 74 & 80 & 95 & 121 & 404 \\
55603.6 & 19.41 & 252 & 65 & 71 & 79 & 94 & 118 & 394 \\
55666.5 & 19.57 & 63 & 71 & 79 & 85 & 102 & 130 & 432 \\
55957.6 & 19.26 & 291 & 60 & 65 & 74 & 85 & 109 & 354\\
\hline
\end{tabular}
\label{CT_ugc_high}
\end{center}
\end{table*}

\begin{table*}[h]
\begin{center}
\caption{Survey epochs and CTs of IRAS 16516-0948.}
\begin{tabular}{ccccccccc}
\hline
\hline
    MJD & $m_{\mathrm{Lim}}$ & $\Delta t$ & IIb & Ib & Ic & II & 87A & IIn \\
    &(mag)& (d)  &  (d)& (d)& (d)& (d)& (d) & (d) \\
\hline
54577.6 & 19.93 & - & 69 & 76 & 84 & 96 & 126 & 417 \\
54724.2 & 18.97 & 147 & 35 & 40 & 47 & 54 & 68 & 191 \\
54928.6 & 18.39 & 204 & 20 & 22 & 28 & 28 & 41 & 98 \\
54957.5 & 19.96 & 29 & 70 & 77 & 84 & 98 & 125 & 428 \\
54986.4 & 19.67 & 29 & 60 & 66 & 73 & 83 & 108 & 348 \\
55307.6 & 19.93 & 321 & 69 & 77 & 85 & 96 & 124 & 418 \\
56141.3 & 19.70 & 834 & 61 & 68 & 75 & 85 & 110 & 360 \\
\hline
\end{tabular}
\label{CT_iras16516_high}
\end{center}
\end{table*}

\begin{table*}[h]
\begin{center}
\caption{Survey epochs and CTs of IRAS 17138-1017.}
\begin{tabular}{ccccccccc}
\hline
\hline
    MJD & $m_{\mathrm{Lim}}$ & $\Delta t$s & IIb & Ib & Ic & II & 87A & IIn \\
    &(mag)& (d)  &  (d)& (d)& (d)& (d)& (d) & (d) \\
\hline
54577.6 & 19.18 & - & 55 & 60 & 68 & 79 & 101 & 323 \\
54613.5 & 19.27 & 36 & 57 & 64 & 72 & 83 & 106 & 340 \\
54642.5 & 19.87 & 29 & 80 & 88 & 95 & 113 & 144 & 498 \\
54729.2 & 19.64 & 87 & 71 & 78 & 86 & 101 & 128 & 434 \\
54928.6 & 18.86 & 199 & 44 & 49 & 56 & 65 & 84 & 248 \\
54956.6 & 19.69 & 28 & 74 & 80 & 88 & 104 & 135 & 453 \\
54986.4 & 19.41 & 30 & 64 & 69 & 77 & 90 & 115 & 371 \\
55310.6 & 19.32 & 324 & 59 & 66 & 73 & 86 & 108 & 356 \\
56142.3 & 19.64 & 832 & 72 & 78 & 86 & 100 & 131 & 437\\
\hline
\end{tabular}
\label{CT_iras17138_high}
\end{center}
\end{table*}

\begin{table*}[h]
\begin{center}
\caption{Survey epochs and CTs of IRAS 17578-0400.}
\begin{tabular}{ccccccccc}
\hline
\hline
    MJD & $m_{\mathrm{Lim}}$ & $\Delta t$s & IIb & Ib & Ic & II & 87A & IIn \\
    &(mag)& (d)  &  (d)& (d)& (d)& (d)& (d) & (d) \\
\hline
54578.6 & 18.49 & - & 34 & 38 & 45 & 52 & 65 & 177 \\
54694.3 & 19.82 & 116 & 80 & 88 & 96 & 113 & 144 & 499 \\
54726.3 & 19.98 & 32 & 88 & 94 & 101 & 120 & 157 & 536 \\
54928.6 & 20.01 & 202 & 87 & 94 & 103 & 122 & 157 & 560 \\
54956.6 & 20.03 & 28 & 89 & 95 & 103 & 124 & 157 & 561 \\
55049.3 & 19.84 & 93 & 80 & 89 & 96 & 113 & 147 & 508 \\
55310.6 & 20.03 & 261 & 89 & 97 & 104 & 123 & 160 & 569 \\
55352.5 & 20.01 & 42 & 87 & 96 & 103 & 122 & 159 & 557 \\
56141.4 & 19.98 & 789 & 87 & 94 & 102 & 120 & 156 & 549\\
\hline
\end{tabular}
\label{CT_iras17578_high}
\end{center}
\end{table*}

\begin{table*}[h]
\begin{center}
\caption{Survey epochs and CTs of MCG+08-11-002.}
\begin{tabular}{ccccccccc}
\hline
\hline
    MJD & $m_{\mathrm{Lim}}$ & $\Delta t$ & IIb & Ib & Ic & II & 87A & IIn \\
    &(mag)& (d)  &  (d)& (d)& (d)& (d)& (d) & (d) \\
\hline
54746.6 & 19.23 & - & 76 & 83 & 91 & 108 & 138 & 473 \\
55114.6 & 19.23 & 368 & 76 & 83 & 90 & 107 & 137 & 472 \\
55171.4 & 19.54 & 57 & 88 & 96 & 103 & 124 & 156 & 558 \\
55251.3 & 18.88 & 80 & 62 & 69 & 76 & 90 & 115 & 373 \\
\hline
\end{tabular}
\label{CT_mcg_high}
\end{center}
\end{table*}

\end{appendix}

\end{document}